\documentclass[conference]{IEEEtran}
\IEEEoverridecommandlockouts
\usepackage{cite}
\usepackage{amsmath,amssymb,amsfonts}
\usepackage{algorithmic}
\usepackage{graphicx}
\usepackage{textcomp}
\usepackage{xcolor}
\usepackage{subfig}
\usepackage{scalerel} 
\usepackage{tikz}
\usepackage{pgfplots}
\usetikzlibrary{arrows}
\usetikzlibrary{shapes}

\newcommand*\circled[1]{\tikz[baseline=(char.base)]{
            \node[shape=circle,draw,inner sep=1pt] (char) {#1};}}
\pgfplotsset{width=2cm,compat=1.8}
\usetikzlibrary{svg.path}
\usetikzlibrary{patterns}
\def\BibTeX{{\rm B\kern-.05em{\sc i\kern-.025em b}\kern-.08em
    T\kern-.1667em\lower.7ex\hbox{E}\kern-.125emX}}
    \definecolor{orcidlogocol}{HTML}{A6CE39}
\tikzset{
  orcidlogo/.pic={
    \fill[orcidlogocol] svg{M256,128c0,70.7-57.3,128-128,128C57.3,256,0,198.7,0,128C0,57.3,57.3,0,128,0C198.7,0,256,57.3,256,128z};
    \fill[white] svg{M86.3,186.2H70.9V79.1h15.4v48.4V186.2z}
                 svg{M108.9,79.1h41.6c39.6,0,57,28.3,57,53.6c0,27.5-21.5,53.6-56.8,53.6h-41.8V79.1z M124.3,172.4h24.5c34.9,0,42.9-26.5,42.9-39.7c0-21.5-13.7-39.7-43.7-39.7h-23.7V172.4z}
                 svg{M88.7,56.8c0,5.5-4.5,10.1-10.1,10.1c-5.6,0-10.1-4.6-10.1-10.1c0-5.6,4.5-10.1,10.1-10.1C84.2,46.7,88.7,51.3,88.7,56.8z};
  }
}

\newcommand\orcidicon[1]{\href{https://orcid.org/#1}{\mbox{\scalerel*{
\begin{tikzpicture}[yscale=-1,transform shape]
\pic{orcidlogo};
\end{tikzpicture}
}{|}}}}

\usepackage{hyperref}

\begin{document}

\title{Learning-driven Zero Trust in Distributed Computing Continuum Systems}

 \author{
 Ilir Murturi \orcidicon{0000-0003-1990-0431}, Praveen~Kumar~Donta \orcidicon{0000-0002-8233-6071}, 
 Victor Casamayor Pujol\orcidicon{0000-0003-2830-8368}, 
 Andrea Morichetta\orcidicon{0000-0003-3765-3067}, and Schahram Dustdar\orcidicon{0000-0001-6872-8821}\\[0.2cm]

 \IEEEauthorblockA{{Distributed Systems Group}, {TU Wien},     Vienna 1040, Austria.\\  \{i.murturi, p.donta, a.morichetta, v.casamayor, dustdar\}@dsg.tuwien.ac.at\\
      }
 }
 \IEEEoverridecommandlockouts
\IEEEpubid{\makebox[\columnwidth]{\hfill} \hspace{\columnsep}\makebox[\columnwidth]{ }}
\maketitle
\IEEEpubidadjcol
\begin{abstract}
Converging Zero Trust (ZT) with learning techniques can solve various operational and security challenges in Distributed Computing Continuum Systems (DCCS). Implementing centralized ZT architecture is seen as unsuitable for the computing continuum (e.g., computing entities with limited connectivity and visibility, etc.). At the same time, implementing decentralized ZT in the computing continuum requires understanding infrastructure limitations and novel approaches to enhance resource access management decisions. To overcome such challenges, we present a novel learning-driven ZT conceptual architecture designed for DCCS. We aim to enhance ZT architecture service quality by incorporating lightweight learning strategies such as Representation Learning (ReL) and distributing ZT components across the computing continuum. The ReL helps to improve the decision-making process by predicting threats or untrusted requests. Through an illustrative example, we show how the learning process detects and blocks the requests, enhances resource access control, and reduces network and computation overheads. Lastly, we discuss the conceptual architecture, processes, and provide a research agenda.


\end{abstract}


\section{Introduction}
\label{intro}
In recent years, the number of Internet of Things (IoT) devices has increased substantially in the computing infrastructure. This manifests in increased demand for real-time applications that are fast, secure, and protect end-users' privacy. This rise of IoT has shifted how we think about the processes involved in managing IoT systems and their executions in computing infrastructures. To effectively manage people, devices, and data involved in these systems, it is necessary to ensure that all of these actors are integrated and protected from various unknown threats. In particular, there is a fundamental need for better security models for protecting different digital assets (i.e., resources, data, services, etc.) with highly dynamic authorization decisions in complex and broad scenarios like the device-edge-cloud computing continuum \cite{dustdar2022distributed}.

Traditional security models focus on securing digital asset groups using perimeter-based security or encryption techniques to ensure that only reliable and authenticated actors may enter a secured domain \cite{syed2022zero}. However, the perimeter-based approach cannot work in these broad scenarios due to facts that {(i) it largely ignores insider threats within an authenticated network}, and {(ii) the notion of the perimeter in the computing continuum is hardly applicable}. In this regard, Zero Trust (ZT) represents an appealing direction with perimeter-less and continuous verification capabilities to ensure that digital assets on the computing continuum are protected against potential threats \cite{rose2020zero}. The key principle of ZT is "\textit{never trust, always verify}"; meaning that all network traffic should be strictly monitored and verified before being allowed to access a network or resource \cite{rose2020zero}. Nevertheless, implementing ZT in the computing continuum infrastructures requires careful planning and further advanced mechanisms to improve the decision-making process. 

In a ZT environment, all access to data and resources is strictly controlled and verified, regardless of whether the access is coming from inside or outside the system network. Monitoring activity in a ZT environment involves various tools and techniques \cite{syed2022zero}. These tools enable us to quickly identify and respond to any suspicious or unauthorized activity. When suspicious activity is detected, it is typically investigated using tools and techniques to determine the cause of the activity and evaluate the potential risks. More specifically, this involves analyzing logs, network traffic, or conducting further investigations to identify the source of the activity and any potential damage that may have been caused. Consequently, the outcome of such analysis results in inappropriate actions taken to address the issue. For instance, this may involve blocking access to the requested data or resource, revoking access privileges, blocking lateral movement, or preventing further damage or unauthorized access \cite{yan2020survey}. 

As per the National Institute of Standards and Technology (NIST), ZT is not necessarily a centralized approach, as it can be implemented in a decentralized manner~\cite{stafford2020zero}. However, implementing decentralized ZT in the computing continuum requires careful planning, understanding infrastructure limitations, and introducing further advanced mechanisms to improve the decision-making process. Several challenges exist in implementing decentralized ZT in the computing continuum, such as:
\begin{itemize}
    \item \textit{Limited resources:} Computing continuum infrastructures are three-tier architectures (i.e., cloud, fog, and edge)\cite{murturi2021decentralized} characterized by heterogeneous and dynamic devices. In the edge-tier, edge devices usually have limited resources in terms of processing power, storage, and networking capabilities. Therefore, enforcing complex security measures for edge devices is a challenging task and not always possible.
   \item \textit{Limited connectivity:} Edge devices are usually roaming devices or static devices located in remote or inaccessible areas. Uncertainty is the reason that makes it difficult to establish and maintain continuously a secure connection to the central network, respectively, to the ZT central engine.
    \item \textit{Limited visibility:} Edge and fog tiers represent distributed devices in the computing continuum. Monitoring and managing edge and fog device security metrics in real-time in a centralized manner can be demanding and not always possible.
  \end{itemize}

Incorporating learning strategies along with ZT in the computing continuum will make the control management model intelligent. With learning models, ZT can predict threat requests and block them before proceeding further. It helps to detect non-person entries (NPEs), and therefore, it reduces the amount of resource allocation or malicious actions \cite{bush2022zero}. Furthermore, learning algorithms are computationally intensive; meaning that, they require more resources or time. Since limitations are concerned, lightweight learning approaches are needed to lower resource usage and produce results quickly. Therefore, this can be accomplished with Representation Learning (ReL). There are several ReL approaches in the literature for computing continuum \cite{donta2022promising} matching the needs of ZT. Nevertheless, this article is the first work targeting learning in ZT architecture for Distributed Computing Continuum Systems (DCCS) \cite{info14030198,casamayor2023distributed, pujol2023edge}. The key objective is to introduce a novel decentralized framework that combines ReL and ZT into one platform with the aim of improving security, providing faster resource access controls to end-users, and reducing network and computing overheads. The major contributions of this paper are as follows:

\begin{itemize}
\item We introduce a novel learning-driven ZT conceptual architecture designed for distributed computing continuum systems. We extend the conceptual ZT framework \cite{stafford2020zero} with two novel components (i) learning and (ii) resource management.

\item {We consider Bayesian network structure learning (BNSL) to learn representations from the historical active logs. Via the learning model, we can predict the likelihood of a given request being authentic or fraudulent.}

\item Lastly, we present potential research directions that can foster novel studies in this field and overcome the current limitations.
\end{itemize}

The remaining sections are structured as follows. Related work is presented in \textit{Section}~\ref{rw}. \textit{Section}~\ref{section2} discusses the advantages of learning-driven in ZT-enabled computing continuum infrastructures. Furthermore, we introduce and explain a novel learning-driven ZT conceptual architecture designed for distributed computing continuum systems. In \textit{Section}~\ref{ragenda} we outline a research agenda. Finally, we conclude our discussion in \textit{Section}~\ref{conclusion}.

\section{Related Work}
\label{rw}

In \cite{fu2018intelligent}, \textit{Fu et al.} show that traditional detection techniques can detect attacks depending on their prior modeling; however, such conventional methods may not be fully effective in catching cyber threats because threats emerge frequently. The proposed intelligent attack detection method is based on long short-term memory recurrent neural networks. Its applicability is demonstrated via experiments that show that it can effectively detect anomalous traffic activity in social networks. Similarly, the relevance and feasibility of Convolution Neural Networks (CNNs) to identify cyber threats in real-time have been shown in \cite{kwon2018empirical}. The authors aimed to discover the impact of the structural depths on the overall performance. Therefore, three simple CNN models are evaluated with different internal depths for network anomaly detection. Another research work emphasizes the feasibility and relevance of using ML techniques to detect various attacks on IoT networks \cite{kotenko2019attack}. More specifically, the authors introduced an architecture of the system that allows the detection of abnormal activity in IoT devices using ML techniques (i.e., Decision Tree, Support Vector Machine, Multilayer Perceptron). In \cite{alzahrani2021attacks}, several ML algorithms (i.e., Logistic Regression, Gaussian Naive Bayes, Multi-layer Perceptron Artificial Neural Networks, Random Forest, and Gradient Boosting classifier) are compared and investigated for detecting different attacks in IoT such as DoS and other malicious activities. 

In \cite{ge2019deep}, \textit{Ge et al.} used deep learning to identify attacks on IoT devices. They specifically extracted features from packet headers and used a feed-forward neural network to detect four types of attacks: DoS, DDoS, reconnaissance, and information theft. \textit{Olivier et al. }\cite{brun2018iot} present the principles and design of a deep learning-based approach for the online detection of network attacks. The paper investigates cybersecurity threats faced by an IoT-connected home environment and presents the principles and design of a learning-based approach for detecting network attacks. 

The research works mentioned above focus on solutions for attack detection, traffic detection, and classification with ML techniques for Software-defined network (SDN) architectures. However, there has not been any research that has developed a solution for implementing ZT security in a distributed manner while taking into account the unique characteristics of computing continuum infrastructures. Lastly, there are several learning algorithms used in the literature for cybersecurity, DoS attacks, IoT networks, etc. But, these approaches are resource and time intensive; while learning strategies in ZT have not been considered. Nevertheless, using lightweight learning strategies such as ReL for ZT helps to make it more adaptive and efficient in decision making such as whether to allow or block a request. 

\section{Learning-driven ZT in Distributed Computing Continuum Systems}
\label{section2}

\begin{figure}[ht]
 \centering
  \includegraphics[width=\columnwidth]{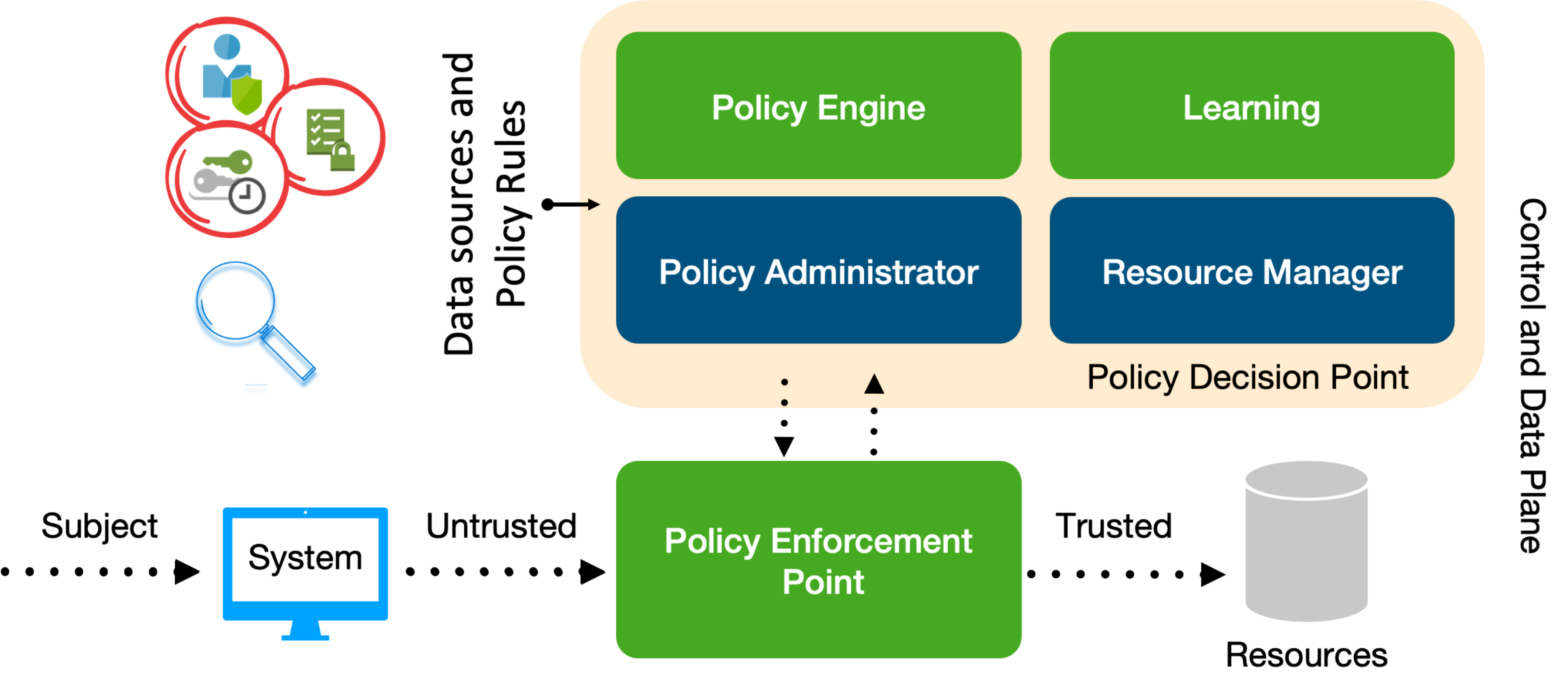}%
 \caption{Zero Trust framework \cite{stafford2020zero} with two additional blocks: \texttt{Learning} and \texttt{Resource Management}.}
  \label{fig1a_ref}
\end{figure}
 
Learning-driven ZT is an innovative approach that comes into play where a set of learning mechanisms in combination with ZT techniques can be applied to secure and increase trustworthiness in the computing continuum. NIST introduced a conceptual framework model of ZT that consists of three main components: \textit{Policy Engine (PE)} is in charge of the ultimate decision to grant access to a resource for a given subject. The engine utilizes information provided by various external data sources (i.e., Activity Logs, ID Management, Data Access Policy, etc.), and through a trust algorithm it grants, denies, or revokes access to the resource; \textit{Policy Administrator (PA)} aims at establishing and/or shutting down the communication path between a subject and a resource; and finally, \textit{Policy Enforcement Point (PEP)} is in charge of monitoring, and terminating connections between a subject and a resource. Three components of the original ZT framework can be observed in Figure \ref{fig1a_ref}. 

However, we advocate the need to expand this framework (as presented in Figure \ref{fig1a_ref}) to be suitable for DCCS. Therefore, we extend the framework and analyze newly added components, explain their implementation challenges, and present promising techniques that support learning model development within ZT. 

\subsection{Learning}
\begin{figure}[ht]
    \centering
    \includegraphics[width=\columnwidth]{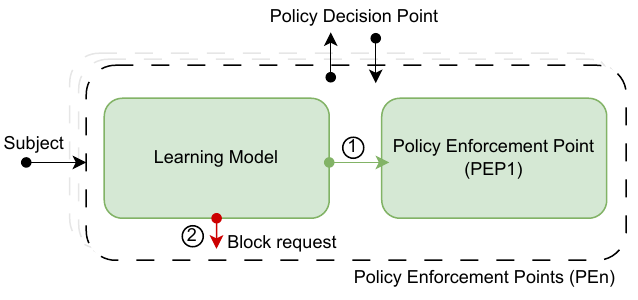}
    \caption{Policy Enforcement Point with learning model.}
    \label{fig_interaction}
\end{figure}

Monitoring and updating the PE in ZT is important to ensure that it is effective at detecting and preventing fraudulent activities for DCCS. Analyzing ZT's activity logs (i.e., from external data sources) helps to detect fraudulent or suspicious requests. In general, representation learning algorithms are useful for identifying the underlying information from data \cite{donta2022promising}, so ZT architecture can benefit from them as well. The ReL algorithms can build a learning model by using different information such as authentication requests and subject behaviors (i.e., user behavior). This learning model can predict a likelihood of a given request being authentic or fraudulent. Accordingly, the decision-making system (PEP1) allows or declines the request. In Figure \ref{fig_interaction}, we show the extended form of the PEP and interactions. First, the learning model continuously evaluates the subject's behavior and determines whether access should be granted \circled{1} or not \circled{2}. If the model determines that subject behavior is trustworthy \circled{1} then the Policy Decision Point (PDP) through PEP decides whether to give access or not to a requested resource. In the case when a connection between PEP and PDP is not stable, the learning model can decide whether access should be granted or not. Notice that PEPs are distributed across the computing continuum to improve overall service quality. Furthermore, we consider Bayesian Network Structure Learning (BNSL) \cite{scanagatta2019survey} to learn from the historical active logs. \textcolor{black}{BNSL due to its probabilistic nature, causal reasoning capabilities, flexibility, interpretability, and ability to handle incremental updates set it apart from other approaches when it comes to learning from historical active logs.}

\begin{figure}[t]
    \centering
    \includegraphics[width=\columnwidth]{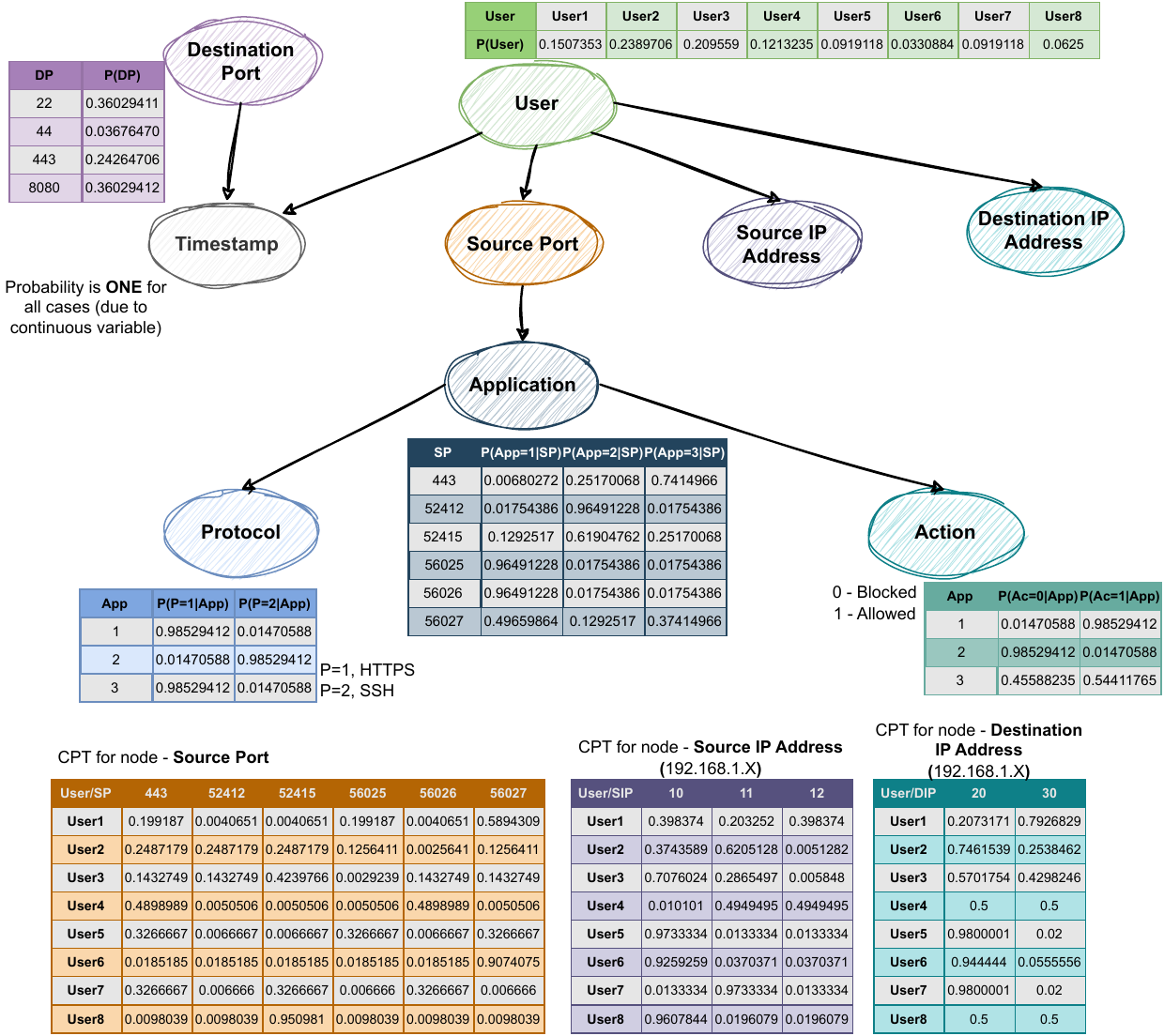}
    \caption{Example structure and conditional probability tables of representation learning through active logs of ZT architecture.}
    \label{fig2}
\end{figure}

In general, an active log can store a huge amount of information. However, we consider a synthetic dataset with attributes (e.g., Timestamp, Source IP Address, Destination IP Address, Source Port, Destination Port, Protocol, User ID, Application, and Action) to evaluate the usability of learning in decision-making for ZT. However, the developers can also consider several other attributes including networks, workloads, visibility and analytics, orchestration, etc. The score-based BNSL analyzed the data and generated a knowledge graph, which looks like a Directed Acyclic Graph (DAG) with weighted links and a Conditional Probability Table (CPT) \cite{rusek2021score}. A weighted link shows mutual information or conditional dependency, and a CPT shows causal uncertainty among attributes. The learned representation and its CPTs are shown in Fig.~\ref{fig2}. We consider a timestamp in an epoch, eight users, three source IP addresses, two destination IP addresses, five source ports, three destination ports, two protocols (SSH and HTTPS), three user applications, and actions (allowed or blocked). We use thirty-three entries (in the illustration example) of active logs to generate Fig.~\ref{fig2}. This learned representation converged with a log-likelihood of \texttt{-284.2246} and Bayesian Information Criterion of \texttt{1113.9043}. 

Using the CPT values illustrated in Fig.~\ref{fig2}, we can also infer whether a new request is allowed or blocked through querying. This querying process can include full or partial information about the request. As this learning process supports missing values, the model infers the results based on available information. For instance, the request is generated with the details of the application, protocol, and source ports are a web browser, HTTPS, and 443, respectively. The learning model queries with this information such as \texttt{P (Action=1| Source Port =443, Protocol ='HTTPS', Application='Web Browser')}. This query returns a probability value for allowed, i.e., \texttt{0.9853} (according to the learned model shown in Figure~\ref{fig2}). Thus, a request with Source Port = 443, Protocol = HTTPS, and Application = 'Web Browser' will be recommended for consideration by the policy engine. In case, a request is generated with (User 8), source IP (192.168.1.10), source port address (56025), and SSH protocol is used. Then, a query \texttt{P(Action=1| Source Port = 56025, Protocol = 'SSH',  User = 'User8', Source IP address = '192.168.1.10')} generates the probability to be \texttt{0.45301}. In such cases, the request is blocked directly, without involving the policy engine. So, the burden on the policy engine is reduced to evaluate the queries. 

\begin{figure}[t]
    \centering
    \includegraphics[width=\columnwidth]{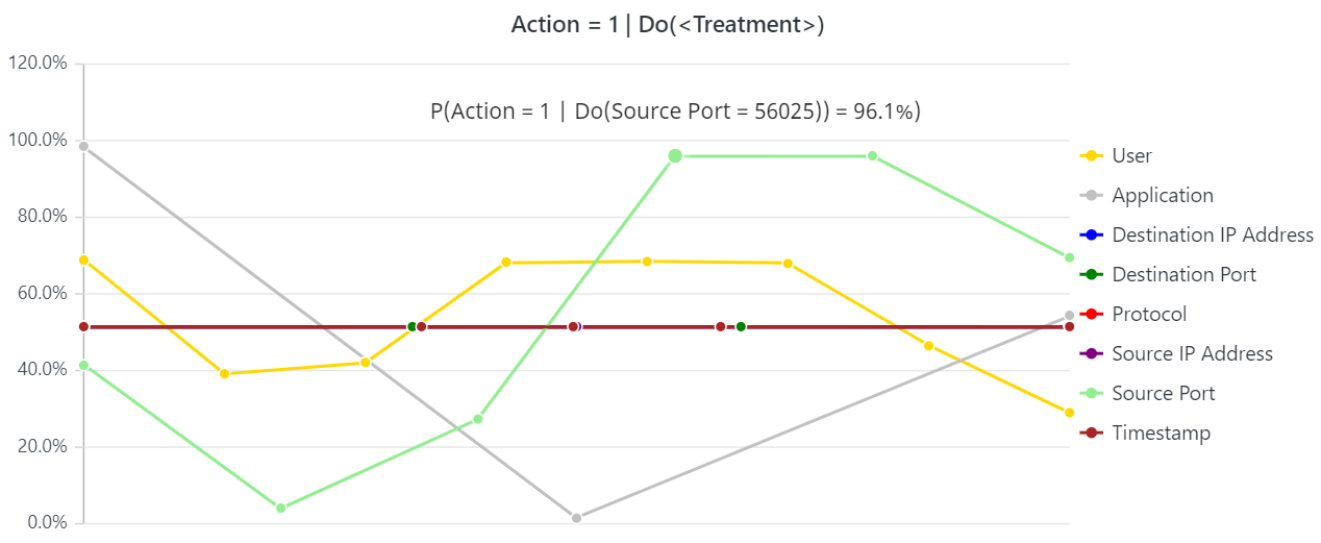}
     \caption{The causal effect analysis on the percentage of \texttt{allowed} request depends on different metrics. }
    \label{fig3}
\end{figure}

Fig.~\ref{fig3} shows the probability of allowing a request depending on each individual attribute through causal effect analysis. In this, the Y-axis shows the probability percentage of allowing, and the X-axis indicates data series of different attributes. Consider the impact of \texttt{Source Port} on \texttt{Action}. The request came from port 56025, creating a probability of 96.1\% that it would be allowed. Similarly, requests from \texttt{Source Port} 52415 resulted in only a 27.4\% probability of allowing the request, which means they might not be trusted. A summary of each metric's effect on action is shown in Fig.~\ref{fig3}. 


\subsection{Resource Management} 
Computing continuum infrastructures are complex and uncertain environments. Implementing a consistent and effective security strategy in these infrastructures is complex and requires a combination of many different technologies and devices. Such challenges are also visible when developing, deploying, and operating applications or systems on such infrastructures. Therefore, we advocate that resource management in ZT is a \textit{must-have} feature since it overcomes several open challenges mentioned previously (see Section \ref{intro}). More specifically, challenges are notable in edge environments where the network perimeter is not clearly defined and can be more easily breached by attackers.  

The resource management component includes several capabilities: 1) discovering available computing resources and monitoring infrastructure to keep track of changes, 
2) self-adaptive monitoring, orchestrating, and determining the appropriate placement for software components in order to provide reliable and low-latency service to end users (i.e., discussed in Section \ref{software}), and 3) distributing and configuring learning models. Nevertheless, the resource management component can be extended with further functionalities. The core resource management component runs only in the cloud, while functionalities (i.e., known as \textit{edge functions} \cite{dustdar2020towards}) can be distributed over the computing continuum depending on resource demands. 

\subsubsection{Self-adaptive and Resilient Runtime Mechanism}
Contrary to cloud infrastructures, computing continuum infrastructures are very heterogeneous environments. 
It is crucial to have a way to execute functionalities on a single, consistent, and lightweight runtime platform that allows them to be executed on any device without the need for additional configuration. As a potential candidate, we consider an open standard technology called WebAssembly\footnote{WebAssembly, https://webassembly.org/}, which provides full interoperability across different platforms. WebAssembly is a lightweight mechanism with several benefits (i.e., low-latency,  dynamic and scaleable, language agnostic, etc.) over other virtualization platforms such as Docker or Java-based OSGi\footnote{Java-based OSGi, https://osgi.org/}. Considering that computing continuum environments are characterized by uncertainty\cite{krupitzer2015survey,weyns2021towards}, we require novel decision-making and an intelligent self-adaptive orchestrator for resource discovery and placement, resource provisioning, and adaptive monitoring, just to name a few. In this paper, we treat resource management aspects as future work and orthogonal to our approach; we are concerned with the core learning mechanisms to show the feasibility to improve the ZT security and reduce network and computations
overhead in computing continuum environments.

\subsubsection{Configuration and Distributing Learning Models}
Computing continuum environments provide a seamless opportunity to train a shared model on multiple devices, such as smartphones, fog, or edge devices. Rather than sharing the raw data, sharing their locally-computed gradients reduces the amount of data that needs to be transmitted between devices and a central server. Optimizing communication between low-powered devices and a centralized server is crucial, especially for saving device energy. {Since BNSL's time complexity is not time intensive \cite{scutari2019learning}, it can run on computing entities across 
computing continuum. Additionally, it minimizes communication delays and ensures that active logs are not misused.} 

\subsection{Deployment Considerations}
\label{software}

\begin{figure}[ht]
 \centering
  \includegraphics[width=\columnwidth]{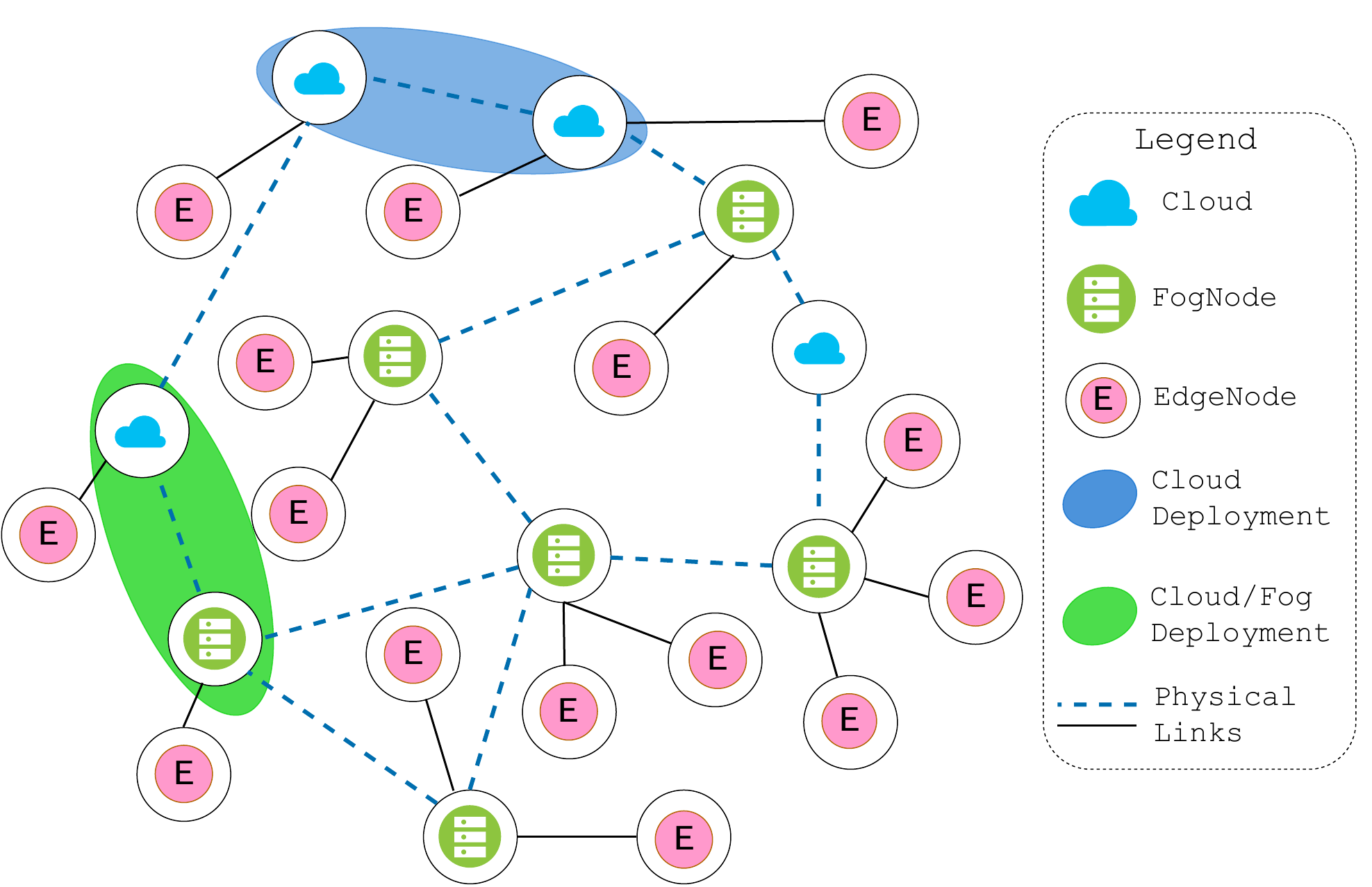}%
 \caption{Software architecture considerations in computing continuum. }
  \label{fig1b}
\end{figure}

Software components including their functionalities (i.e., edge functions), and learning shared models can be placed on different devices in the computing continuum, yielding different deployments and configurations. As depicted in Figure \ref{fig1b}, deployment types can be categorized into two models: (1) cloud model (i.e., depicted with blue color), and (2) edge-cloud hybrid model (i.e., depicted with green color). As can be noted in Figure \ref{fig1b}, software components such as Policy Administrator and Resource Management can be deployed only in the cloud (i.e., however, some functionalities can be distributed across the computing continuum). These components are expected to provide core functionalities such as orchestration and storage; therefore, having them in the cloud is essential due to the high availability of resources. The other three software components or their features are distributable across the computing continuum. For example, software components that receive many requests from a particular region can be placed close to end-users to improve performance and overall latency. Similarly, a shared learning model is distributed across the entire continuum, and then each device trains the model locally on its data.

\section{An Emerging Research Agenda} 
\label{ragenda}

The ultimate goal of learning-driven ZT is to improve decision-making convenience and enhance the security in computing continuum networks and systems by using representation learning techniques. However, using learning approaches and orchestrating architecture components in dynamic environments does entail several challenges. We identify {four} research directions that must be further investigated in the future:

\subsection{Resource management in computing continuum} Resources are distributed across different network layers in the computing continuum, ranging from low-powered edge devices to cloud servers. Therefore, emerging paradigms such as multi-domain orchestration, the orchestration of components based on WebAssembly, or Function as a Service (FaaS) orchestration are becoming increasingly prevalent in the computing continuum. However, current implementations lack mechanisms for distributed orchestration and decision-making responsibilities across the computing continuum (i.e., these existing paradigms are often centralized). Therefore, this is challenging because fully centralized approaches are often unsuitable for computing continuum systems.

\subsection{Parameter and model selection} In learning-driven zero trust,  selecting appropriate security parameters is crucial to ensure the effectiveness and efficiency of the system. In general, zero trust relies on collecting and analyzing data from multiple sources, such as network traffic, user behavior, and system logs. However, selecting specific parameters will depend on the types of data sources available, the quality of the data, and the computational resources required to process it. More specifically, choosing specific parameters will vary based on the needs and requirements of each enterprise. Therefore,  selecting the most appropriate learning models for a given task requires understanding the problem domain, the available data, and the computational resources available.

\subsection{Incremental learning in ZT} DCCS must continuously adapt to changing threats and vulnerabilities. The learning model must constantly learn from new data without forgetting previous knowledge and enhance adaptability. This process keeps the learning model to be up to date and able to detect new threats. Incremental learning further enables scalability which helps to avoid manual updates or reconfiguration. Therefore, developing novel methods and algorithms for continuous learning in ZT-based systems becomes a critical challenge. Such algorithms need to be developed to handle failures gracefully, recover from them, and ensure uninterrupted learning. Moreover, learning becomes even more challenging in distributed environments where data is distributed across multiple sources or devices. In such distributed environments,  consistent and up-to-date models across distributed devices, communication, and synchronization mechanisms are essential.

\subsection{Light-weight AI/ML} In DCCS, resource-constrained edge devices play a major role in computations with low latency. Light-weight AI/ML algorithms that minimize resource usage and time spent computing without affecting prediction accuracy are needed for these edge devices. ML model compression, which reduces redundant data in the models, is one way of achieving lightweight learning models. However, novel methods for lightweight AI/ML in ZT are a challenging issue and such methods should strike a balance between compression and accuracy trade-offs, ensuring that the compressed models still perform well in ZT systems. 

\subsection{Risks and Challenges}
Introducing ReL in ZT adds complexity and may present potential risks.  ReL can be a powerful approach; however, it requires careful design and development to ensure security and effectiveness within a ZT framework. Therefore, addressing various concerns and defending against potential security holes is important. For instance, several measures can be taken, such as ensuring that the data used to train the models is high quality and properly curated. Or there should be clear boundaries for what the learning model can and cannot do to avoid unexpected behavior and potential security holes. Furthermore,  real-time monitoring helps to detect unexpected behavior or anomalies that may indicate security breaches or inaccuracies in learning. Nevertheless, we advocate the benefits that learning brings into ZT, while the mentioned risks and several other aspects should be carefully addressed before using these approaches.

\section{Conclusion} 
\label{conclusion}
Incorporating ReL in ZT helps to improve the decision-making process by predicting threats and untrusted requests before being processed by PDP. The ZT decision mechanism (i.e., PDP) combined with learning models enable distributed decisions on whether to grant or deny access to resources. To achieve such goals, we presented a novel learning-driven ZT architecture to improve the decision-making process for accessing resources in distributed computing continuum systems. Nevertheless, this paper is only a small step toward framework operationalization. In future work, we aim to provide a complete technical framework, including technical and architectural aspects.

\section*{{Acknowledgment}}
Research has partially received funding from grant agreement No. 101079214 (AIoTwin) and by EU Horizon Framework grant agreement 101070186 (TEADAL).

\bibliographystyle{ieeetr}
\bibliography{sample-base} 
\end{document}